\documentstyle[12pt,a4,epsbox,epsf]{article}

\begin{document}

\sloppy

\def\thefootnote{\fnsymbol{footnote}}

\begin{center}

{\huge Self-organized hierarchical structure\\ in a plastic
network of chaotic units}

\vspace{1cm}

{\Large Junji Ito\footnote{Corresponding author; e-mail: ito@complex.c.u-tokyo.ac.jp}, Kunihiko Kaneko}

\vspace{0.5cm}

{\it Department of Pure and Applied Sciences, College of Arts and Sciences, 
University of Tokyo,\\ Komaba, Meguro-ku, Tokyo 153, Japan}

\end{center}

\vspace{1cm}

\begin{abstract}
Formation of a layered structure is studied
in a globally coupled map of chaotic units
with a plastic coupling strength 
that changes depending on the states of units globally
and an external input.
In the parameter region characterized 
by weakly chaotic and desynchronized dynamics,
units spontaneously form a hierarchical structure 
due to the influence of the input.
This hierarchical structure is not fixed in time, 
and is successively reorganized.
It is found 
that the distribution of lifetimes of the structure obeys a power law.   
The possible relevance of the present result 
to information processing in the brain 
is briefly discussed.
%\end{abstract}

\begin{center}
{\bf Keywords}
\end{center}

Chaos,
Self-organization,
Globally coupled map,
Flexible coupling, 
Hierarchical structure, 
Power law

\end{abstract}

\section{Introduction}

Many neurophysiological experiments have revealed that the cerebral 
cortex splits into several functionally specialized areas.
In particular, this has been studied in detail 
for the visual cortex \cite{van Essen}.  
The specialized areas form a layered structure 
connected hierarchically and reciprocally.  
Visual information from external inputs is processed 
within this hierarchical structure, 
which constitute of about 30 visual areas.

Although most of such global cortical structure is thought 
to be genetically determined, 
some parts are plastic and organized dynamically 
in a manner that depends on the external inputs 
\cite{Blakemore, Roe, Merzenich}.
In order to determine whether genetic information is necessary
for the formation of this structure, it is important to study 
if such hierarchical structure can be formed spontaneously 
through dynamics alone.

In related studies, based on both theoretical and experimental results,
fast synaptic change has recently been proposed 
as an important process 
in the temporal coding of information \cite{Malsburg, Tsodyks}.
If hierarchical structure is relevant to information processing, 
it is important to study 
if plastic change in connections can lead to dynamic hierarchical structure, 
depending on external inputs, even without pre-determined instructions.
In the present Letter, 
we give an example of the dynamic organization of such structure 
to process external inputs. 

As an abstract model for neural information processing, 
we adopt a simple dynamical system model 
that has the following three properties:
(1) global connections between nonlinear units 
that can exhibit chaotic and other dynamic behavior, 
(2) plastic change of the connections between units
that depend not only on the states of the two units to which they correspond
but also globally on all the other units, 
and (3) an external input applied to one or a few units to change their states.
With this model, we will show 
how a hierarchical structure with several layers is formed spontaneously 
and how this formation depends on the external input.

\section{Model}

To consider the problem described in \S 1, 
we employ a globally coupled map (GCM) \cite{Kaneko} with plastic couplings. 
In a GCM, nonlinear units interact globally.  
Although couplings between units are fixed in a conventional GCM,
we need to introduce a process 
corresponding to the change of the synaptic weight.
For this purpose, here we consider 
variable connection strengths between units, 
which are functions not only of the internal states of the two units 
to which they correspond but also of all other units.
Another missing feature in a conventional GCM 
as a model for a neural system is the response to external inputs, 
which is also included here.

In some applications of GCMs to neural networks,
a unit of the GCM corresponds to a single neuron \cite{Nozawa, Ishii}.
Here we do not adopt such a representation.
Instead we study neural dynamics at a coarse-grained level,
by regarding the dynamics of each unit of the GCM 
as a collective variable representing an ensemble of neurons.
Of course, there is no standard equation accepted as a generally valid model
for such collective behavior,
although it can be represented by a one dimensional iteration map for
some neural network models \cite{Tsuda, Bauer}.
Here, we are interested in the universal behavior of systems 
possessing the features described above, 
rather than the phenomena specific 
to any particular choice of the unit dynamics.  
We wish to introduce a general manner of thinking 
with regard to the phenomena in question, rather than a particular model.

For the units of the GCM to be considered, we adopt the circle map,
\begin{equation}
x_{n+1} = x_{n} + \Omega + \frac{k}{2\pi} \sin 2 \pi x_{n},
\label{eq.unit}
\end{equation}
for the phase variable $x_n$ at the time step $n$.
This map is chosen because it displays oscillatory and chaotic dynamics.  
Indeed, the phenomena to be discussed are universal 
for a collection of units displaying such dynamics.
With units of this type, 
we adopt the following GCM model with plastic couplings:
\begin{equation}
x^{i}_{n+1} = x^{i}_{n} + \Omega + \frac{k}{2\pi} \sin 2 \pi x^{i}_{n} + \frac{c}{2\pi} \sum_{j=1}^{N} \varepsilon^{ij}_{n} \sin 2 \pi x^{j}_{n} + I^{i}.
\label{eq.GCM}
\end{equation}
Here $x^{i}_n$ is the phase of the unit $i$ at the time step $n$,
and $N$ is the number of units.  
The variable $\varepsilon^{ij}_n$ is time-dependent strength 
of the coupling from unit $j$ to unit $i$,
and $I^{i}$ is an input to the unit $i$, 
which works as a driving force to rotate the phase.  
Here we consider only the case of a constant input 
to a single unit or a small number of units.
$\Omega$, $k$, and $c$ are parameters.

In general for systems of the type we study, 
the variation of the connections has been described 
by Hebbian-type dynamics that, 
for a given connection, depend only on the states of the two units 
to which this connection corresponds, 
while other mechanisms have been introduced 
to effect some change on the connections depending globally on other units 
\cite{Edelman,Tsukada}.

It is not straightforward 
to introduce the coupling dynamics in the present case, 
since the state $x^i_n$ refers not to a neuron but to a collective state.
Here we extract the essence of the above dynamics of synapses,  
and include it in our abstract GCM model.  
First, a two-unit Hebbian-type dynamics is introduced, 
to increase or decrease the connection weight,
depending on the degree of synchronization between the two units in question.
Second, the global effect is introduced as a competition 
for the connections converging on a single unit.  
As an example, we consider the dynamics described by 
\begin{equation}
\varepsilon^{ij}_{n} = \frac{\tilde{\varepsilon}^{ij}_{n}}{ \sum_{j=1}^{N} \tilde{\varepsilon}^{ij}_{n}},
\label{eq.normalize}
\end{equation}
\begin{equation}
\tilde{\varepsilon}^{ij}_{n+1} = \cases{ 0 & (for i=j) \cr [1+\delta \cos 2 \pi (x^{j} - x^{i})] \varepsilon^{ij}_{n} & (for i$\ne$j) \cr }.
\label{eq.Hebbian}
\end{equation}
\[(\delta :parameter) \]
The `normalization' over all units in Eq.(\ref{eq.normalize}) 
gives a simple representation of the global competition 
for the coupling change \cite{Kaneko2}.  

While most of the numerical results we present were obtained with 
$\Omega = 0$ and $\delta = 0.1$, 
the conclusion to be drawn is independent of this specific choice.
As an initial condition, 
all the coupling strengths $\varepsilon^{ij}$ were set to be identical,
and the phase variables $x^{i}_0$ was chosen from a random number 
between 0 and 1.

\section{Results}

First we briefly discuss the behavior of the GCM with fixed 
identical couplings, without inputs. 
In general, there are a number of phases in the GCM,
as determined by the two characteristic parameters, 
one governing the nonlinearity of each unit, 
and the other representing the strength of connections between units.
As these two parameters are changed, 
the following four phases are generally observed \cite{Kaneko}:
(i) a coherent phase, where all units oscillate synchronously;
(ii) an ordered phase, where units split into a few clusters 
in which the units oscillate synchronously;
(iii) a partially ordered phase, 
consisting of both synchronized clusters and desynchronized units;
and 
(iv) a desynchronized phase, without synchronization between any two units.
These four phases again appear 
as the parameters $k$ and $c$ are altered, 
for the present model with fixed identical couplings.

In the case that the connection strength changes according to 
Eqs.(\ref{eq.normalize}) and (\ref{eq.Hebbian}), 
the partially ordered phase does not appears.
Figure 1 is the phase diagram for this system,
in which inputs are not yet included.

Now we consider what happens 
when we introduce an external input to one of the $N$ units. 
As we have found, in this case the phase diagram is almost unchanged 
from the case without inputs, depicted in Fig.1.
However, in the present case, 
there appears difference in the correlation between units, 
which results in an inhomogeneity and some non-trivial structure 
in the connection matrix $\varepsilon^n_{ij}$.
One characteristic measure for the effect of an input
is the average variation of $\varepsilon_{ij}$ per step, 
$| \langle \varepsilon^{n+1}_{ij}-\varepsilon^n_{ij} \rangle |$,
where $\langle \cdots \rangle$ is the temporal average. 
We have computed the difference between these variations 
for the cases with and without inputs.  
This difference is plotted in Fig.2
with respect to $k$ and $c$.
As shown in the figure, a strong input dependency is observed 
in and near the region satisfying $3.5<k<4.2$ and $0.2<c<2$, 
located in the desynchronized phase.

\begin{center}
\begin{minipage}{6cm}
\epsfile{file=phase.eps,width=6cm}
{\scriptsize Figure 1: Phase diagram of the GCM described by Eq.(2) 
in terms of the parameters $k$ and $c$, 
with the connection strength defined through Eqs.(3) and (4). 
Here we have $\Omega = 0$ and $\delta = 0.1$, with the system size $N=10$. 
This diagram is obtained without an external input, 
but no changes in the phase diagram were observed 
when we applied an input to a single unit.}
\label{phase}
\end{minipage}
\end{center}

A snapshot of the connection matrix in this regime 
is plotted in Fig.3, where the magnitude of $\varepsilon_{ij}$ 
is represented by the size of the solid square located in the i,j-th grid site.
As seen in this figure, connections with large coupling values are
concentrated in the leftmost column.
Thus we see that the connections from the unit 
to which the external input is added are the most strongly enhanced.

As a visualization method to detect the structure 
existing among the connection strengths, 
we introduce a threshold value for the connection strength 
and regard there to be a connection between the units $i$ and $j$
if and only if $\varepsilon_{ij}$ is greater than this threshold. 
The structure extracted using this method with a threshold value of 0.1
is illustrated in Fig.4,

In this figure, 
the unit to which the external input is applied is shown in the leftmost site,
and units which have connections from this unit 
are located at the next layer,
and units which have connections from these units 
are located at the next right layer, and so forth.  
Although the tree structure from layer to layer 
might look like an artifact of our specific method of plotting, 
this is not the case.
Indeed, with this method, a characteristic feature is revealed: 
The coupling strength to the neighboring layer is much stronger than
that to a distant layer.
Numbering these layers in order of appearance from left to right, 
it is found that almost all units in the $n$-th layer connect 
only with units in either of the $(n-1)$-th, $n$-th, and $(n+1)$-th layer.  
In this sense, it is seen that 
the structure here is not an artifact of our visualization, 
but in fact characterizes inherent structure of the model.  
From this point, we refer to this type of ordered structure as `layered',
and define the layers with small $n$ to be `upper' layer.

\begin{center}
\begin{minipage}{8cm}
\epsfile{file=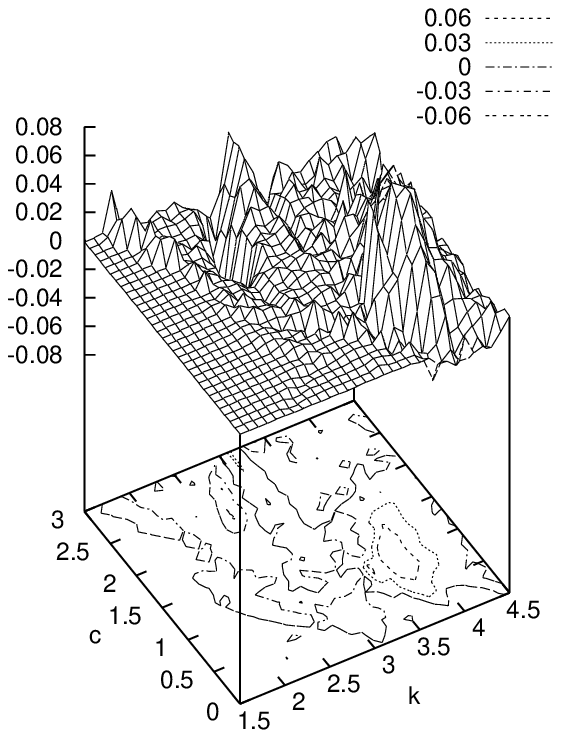,width=8cm}
{\scriptsize Figure 2: Plot of the difference 
between the variations of $\varepsilon_{ij}$ with and without input. 
In and near the parameter region satisfying $3.5<k<4.2$ and $0.2<c<2$, 
the difference is large.  
This region is located in the desynchronized phase, shown in Fig.1.}
\label{sensitivity}
\end{minipage}
\hspace{1cm}
\begin{minipage}{6cm}
\epsfile{file=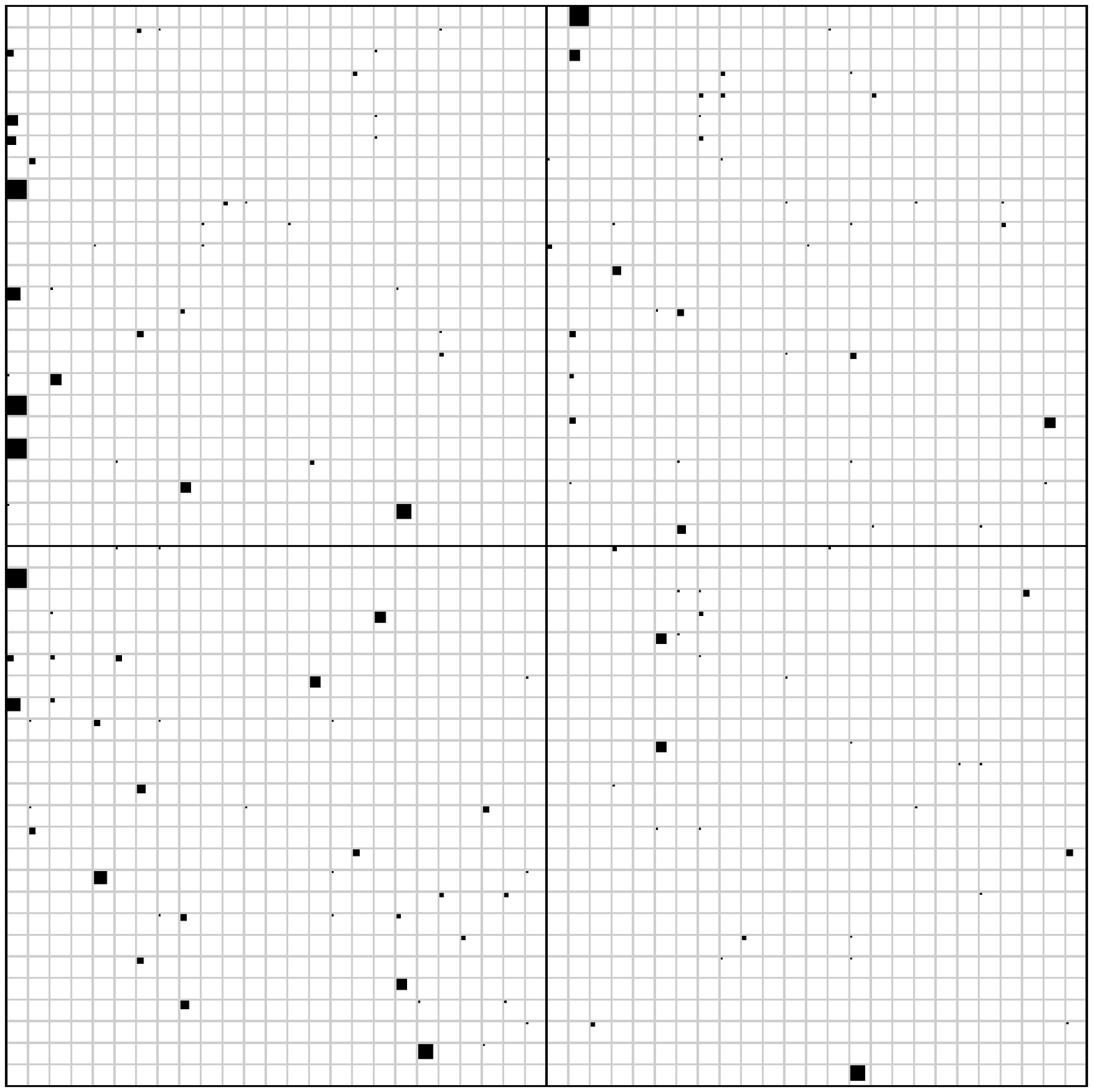,width=6cm}
{\scriptsize Figure 3: Snapshot of the connection matrix. 
The size of the square plotted in i-th row and j-th column 
is proportional to the value of $\varepsilon_{ij}$.  
Here $\varepsilon_{ij}=1$ when the square size is equal to the grid size. 
In the case shown here 
an external input with $\Omega = 0.1$ is applied to the first unit.  
We chose the parameter values $k=4.1$, $c=1.0$, and $\delta = 0.1$, 
which were also used in the computations depicted in subsequent figures.}
\label{matrix}
\end{minipage}
\end{center}

Of course, when the threshold value is varied,
the structure of the connection matrix changes.
However, although the total number of connections 
depends on the value of the threshold, 
the general nature of the results we obtain 
is the same for a significant range of threshold values 
(approximately .02-.3).

\begin{center}
\begin{minipage}{10cm}
\epsfile{file=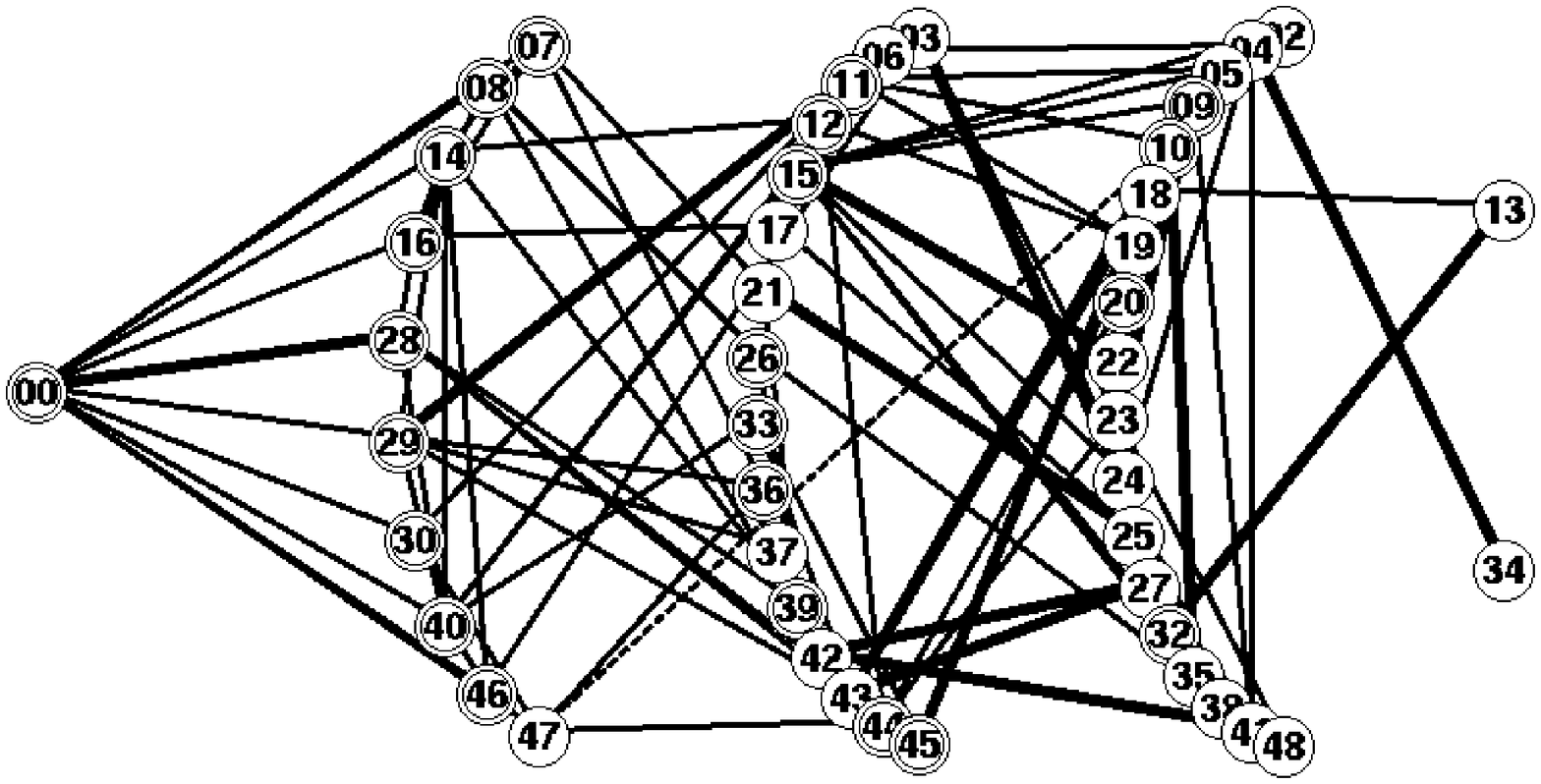,width=12cm}
{\scriptsize Figure 4: Snapshot of the layered structure of the connection, 
for the case that the threshold value is 0.1. 
A thin line indicates a connection from left to right unit, 
and a thick line indicates that from right to left. 
Note that there is only a single connection between units 
that are farther from each other than one layer.
This connection is represented with the broken line in the figure.}
\label{graph}
\end{minipage}
\end{center}

With the dominant presence of nearest-neighbor connections 
in this layered structure, 
we call a connection between neighboring layer 
a `layered structural connection' (LSC), 
while a connection between units separated by more than one layer is called
a `non-layered structural connection' (NLSC). 
The number of LSC indicates the size of the structure,
while the number of NLSC represents ``ambiguit'' of the layered structure.
To determine if such layered structure is generic to a system with an input, 
we have plotted the numbers of LSC and NLSC 
as functions of time for a particular system.   
These plots appear in Fig.5, 
where the external input is applied 
only between the 10,000th and 30,000th steps, 
while for the first 10,000 and the last 20,000 
({\it i.e.} 30,000-50,000) steps, no external input is added.
\begin{center}
\begin{minipage}{8cm}
\epsfile{file=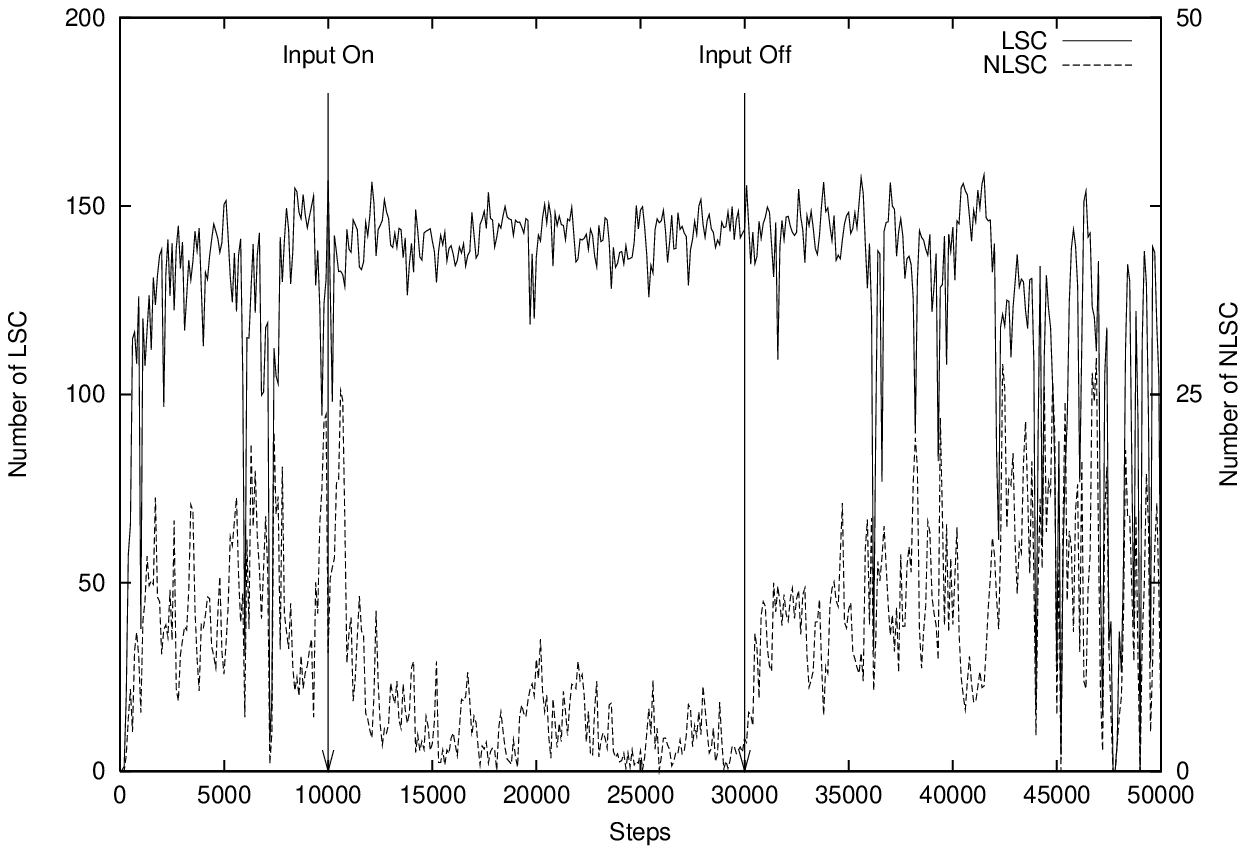,width=8cm}
{\scriptsize Figure 5: Plot of the numbers of LSC and NLSC 
as functions of time. 
The average numbers over intervals 100 steps are plotted.  
The external input is applied only between the 10,000th and 30,000th steps. 
Different scales are adopted for the LSC and NLSC 
to show the input dependence clearly.}
\label{connection}
\end{minipage}
\end{center}

Even during the first 10,000 steps, without inputs, 
the number of LSC is often large.  
However, there are several drops in this number,
indicating that the structure easily collapses.
On the other hand, the number of NLSC is not small, 
implying that the structure is not clearly layered.
With the introduction of the external input at the 10,000th step, 
such drops in the number of LSC disappear, 
and simultaneously, the number of NLSC significantly decreases,
implying that a layered structure with long-term stability appears.
Then, after the input is cut off at the 30,000th step, 
the number of LSC again exhibits several drops,
while the number of  NLSC increases.

Now we show that the layered structure is maintained 
by the introduction of the external input.  
Note that the formation of this layered structure is possible 
only in the parameter regime with sensitive input dependence.  
In other words, organization of structure is possible only
from a weakly desynchronized state.
For an ordered state, the clustered structure is already rigid, 
and there is no freedom to form an input-dependent structure, 
while for a fully desynchronized state, the layered structure,
even if formed, decays rapidly.

Due to the continuous change of the connection matrix, 
the layered structure is not fixed in time; 
{\it i.e.} the units belonging to each layer change in time.
Generally speaking, the upper layer is more stable than the
lower layer.  To check this stability quantitatively, 
we define an `average layer' for each unit as follows:
Let $l_{n}$ be the number of the layer 
to which a given unit belongs at time step $n$.
Then, the average layer for this unit over given $T$ time steps
is defined as $\Sigma_{n=1}^{T}l_{n}/T$.
Then, the frequency of layer switching of a unit is defined 
as the number of changes in layers that that unit experiences 
over the $T$ time steps.
In Fig.6 the frequency per step is plotted with respect to the average layer.
The increase of stability at upper layers is discernible.
\begin{center}
\begin{minipage}{7cm}
\epsfile{file=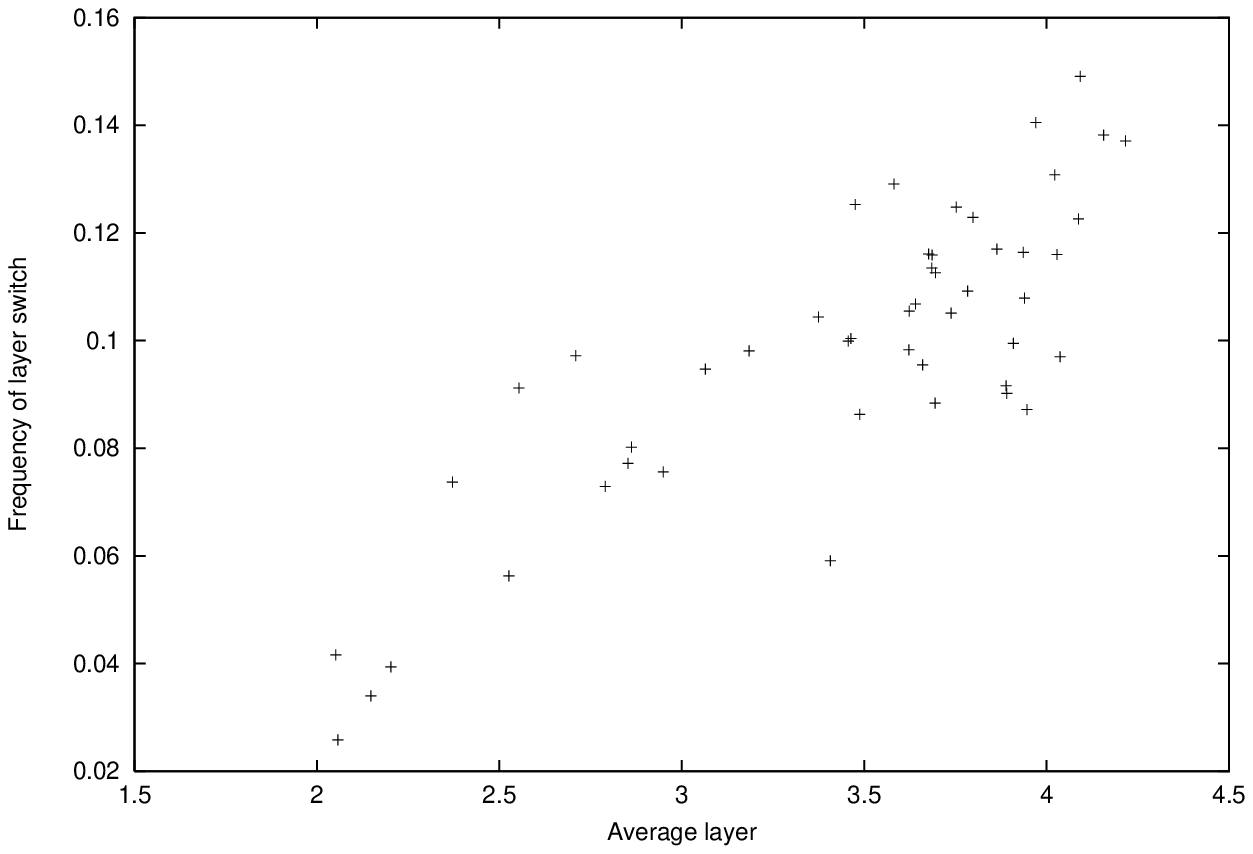,width=7cm}
{\scriptsize Figure 6: The frequency of layer switching 
plotted against the average layer, for each unit.  
(See text for details.) }
\label{layerchange}
\end{minipage}
\hspace{1cm}
\begin{minipage}{7cm}
\epsfile{file=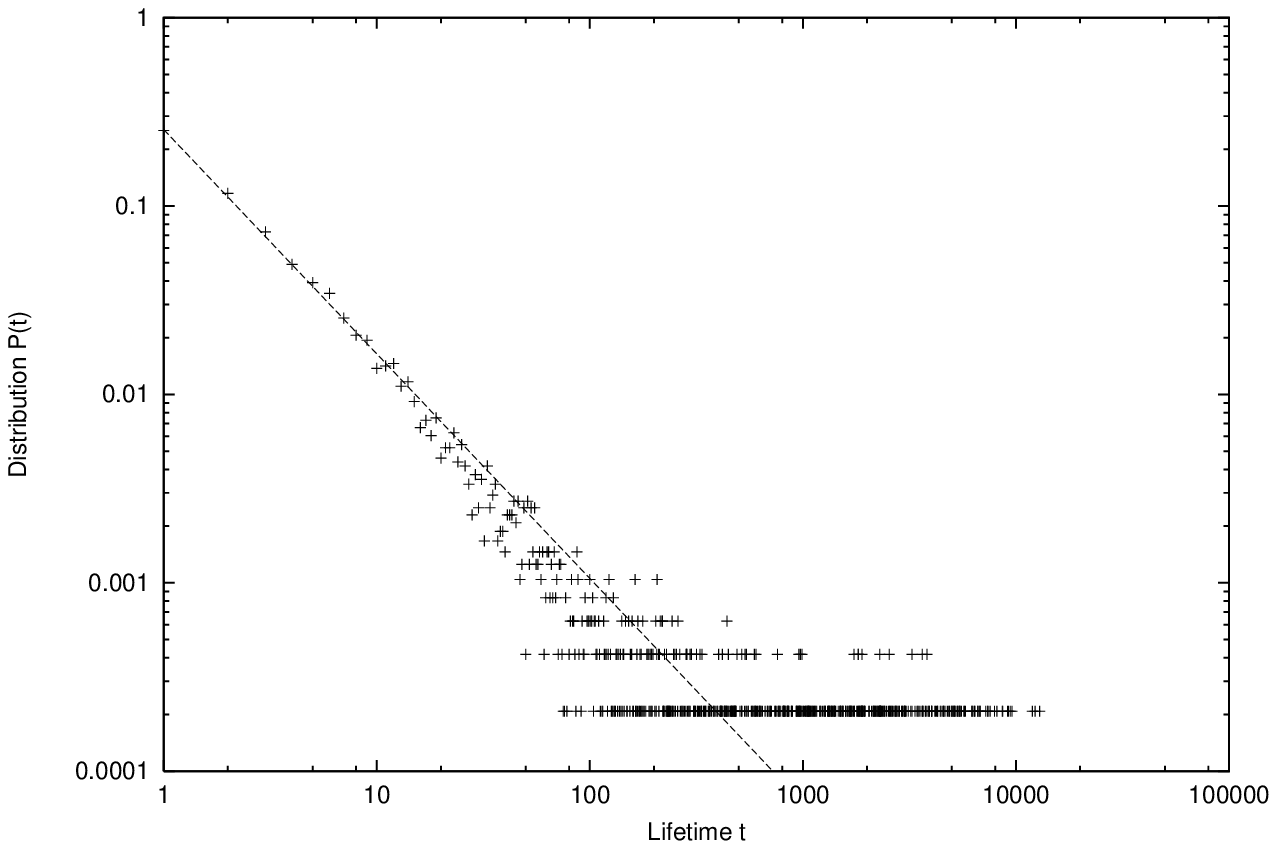,width=7cm}
{\scriptsize Figure 7: Log-log plot of the distribution $P(t)$ 
of lifetimes $t$ of units at the second layer. 
The broken line gives $P(t)\sim t^{-1.2}$, obtained by the least square fit.}
\label{lifetime}
\end{minipage}
\end{center}

The above result implies that the lifetime for which a unit stays
at a given layer is larger for the upper layer.  
Next, we study the distribution of lifetimes of the units at each layer.   
Figure 7 displays the distribution of lifetimes at the 2nd layer 
plotted on a logarithmic scale.
As shown, the distribution obeys a power law, with the exponent about 1.2.  
Interestingly, this power law with the same exponent is observed 
in every layer, although this power law behavior extends
to longer times for the upper layers.

It is important to note that the presently considered self-organization
of hierarchical structure apparently exists for any sufficiently large $N$.
With an increase in size, the number of layers increases, 
and the time scale for the formation and collapse of the layered structure 
also increases.

\section{Discussion}

To sum up, we have found the spontaneous formation of hierarchical structure
in a GCM with a plastic coupling and an input.  
This formation is possible within the parameter region
characterized by a sensitive dependence on the input.
Here, the change of connection matrix is strongly affected by the input.
In other words, 
the spontaneous formation of layered structure is made possible 
by choosing parameter values that allow for high plasticity.

Even in the partially ordered phase of a GCM (with a fixed coupling), 
hierarchical clustering is found \cite{Kaneko},
in which units split hierarchically into clusters
with respect to the synchronization of oscillators.
Such potentiality to form hierarchy is revealed in the present system
by adding an input to one unit. 
Here, the hierarchy is represented by the layered structure 
in the connection matrix.  
In the regime in which this structure appears, 
each unit displays chaotic dynamics, 
without any clear synchronization between any two units. 
In spite of this desynchronization, a layered structure is formed, 
which may suggest the existence of some kind of hidden order \cite{Kaneko3} 
in the motion.

When units form synchronized clusters (in the ordered phase), 
the correlation between two units is fixed.  
The connection between two synchronized units is strengthened,
and a rigid clustered structure among the connections is formed.  
Hence, in this case there is no freedom for the system 
to form a layered structure induced by the input.
On the other hand, if the chaotic behavior of each unit is too strong, 
the correlation in the oscillation between two units is too weak
for any structure in the connection matrix to form.
The formation of hierarchical structure is allowed 
only in the parameter regime characterized by weakly chaotic dynamics, 
where both a rather rigid structure at upper layers 
and unstable change at lower layers coexist.

The presently studied mechanism to form hierarchical structure works 
generally for a globally coupled system of chaotic maps 
(e.g., the logistic map),
as long as we choose parameter values that result in weak chaos.  
However, it seems that the dynamics of the coupling strengths 
must be of a global nature.
Hierarchical structure has not yet been observed
in the case of Hebbian dynamics only.
By the normalization in Eq.(3) the degree of synchronization,
not only between the two units to which a given connection correspond, 
but relatively to other units, 
is employed as the force to strengthen the connection.
Use of such relative synchronization seems to be necessary 
for the system to form hierarchical structure.
It appears that the global competition to enhance the coupling is essential 
for the spontaneous formation of the structure.

The organized hierarchical structure we have found is not static, 
but rather changes in time.
The rapid change at lower layers can influence upper layers from time to time,
leading to the collapse of the layered structure.
The lifetime of the structure is found to obey a power law distribution.

The inherent tendency to form a layered structure that we have observed
may provide a novel viewpoint on the formation of such structure in the brain, 
as well as the possibility of hierarchical temporal information processing 
on short time scales.

\vspace{1cm}
\begin{flushleft}
{\Large \bf Acknowledgments}
\end{flushleft}
\vspace{0.5cm}

The authors would like to thank S. Sasa and T. Ikegami for their valuable comments.
This research was supported by
Grant-in-Aids for Scientific
Research from the Ministry of Education, Science, and Culture
of Japan.

\end{document}